\documentclass[aps,prl,twocolumn,10pt,amsmath,amssymb,
superscriptaddress,longbibliography,floatfix]{revtex4-2}

\usepackage{graphicx}
\usepackage{dcolumn}
\usepackage{bm}
\usepackage{amsmath}
\usepackage{amssymb}
\usepackage{booktabs}
\usepackage{multirow}
\usepackage{tabularx}
\usepackage{siunitx}
\usepackage[colorlinks=true,citecolor=blue,linkcolor=blue]{hyperref}

\graphicspath{{./Figures/}}

\begin{document}

\title{Ferroelectric switching of interfacial dipoles in \texorpdfstring{$\alpha$-RuCl$_3$}{alpha-RuCl3}/graphene heterostructure}

\author{Soyun Kim}
\affiliation{Department of Physics and Chemistry, Daegu Gyeongbuk Institute of Science and Technology (DGIST), Daegu 42988, Republic of Korea}

\author{Jo Hyun Yun}
\affiliation{Department of Physics, Pohang University of Science and Technology, Pohang, 37673, Republic of Korea}
\affiliation{Center for Artificial Low Dimensional Electronic Systems, Institute for Basic Science, Pohang 37673, Korea}

\author{Junsik Choe}
\affiliation{Department of Physics and Chemistry, Daegu Gyeongbuk Institute of Science and Technology (DGIST), Daegu 42988, Republic of Korea}

\author{Dohun Kim}
\affiliation{Department of Physics and Chemistry, Daegu Gyeongbuk Institute of Science and Technology (DGIST), Daegu 42988, Republic of Korea}

\author{Takashi Taniguchi}
\affiliation{Research Center for Materials Nanoarchitectonics, National Institute for Materials Science, 1-1 Namiki, Tsukuba 305-0044, Japan}

\author{Kenji Watanabe}
\affiliation{Research Center for Electronic and Optical Materials, National Institute for Materials Science, 1-1 Namiki, Tsukuba 305-0044, Japan}

\author{Joseph Falson}
\affiliation{Department of Applied Physics and Materials Science, California Institute of Technology, Pasadena, California, 91125, USA}
\affiliation{Institute for Quantum Information and Matter, California Institute of Technology, Pasadena, California, 91125, USA}

\author{Jun Sung Kim}
\affiliation{Department of Physics, Pohang University of Science and Technology, Pohang, 37673, Republic of Korea}
\affiliation{Center for Artificial Low Dimensional Electronic Systems, Institute for Basic Science, Pohang 37673, Korea}

\author{Kyung-Hwan Jin}
\affiliation{Department of Physics and Research Institute of Materials and Energy Sciences, Jeonbuk National University, Jeonju, 54896, Republic of Korea}

\author{Gil Young Cho}
\thanks{Corresponding author. gilyoungcho@kaist.ac.kr}
\affiliation{Center for Artificial Low Dimensional Electronic Systems, Institute for Basic Science, Pohang 37673, Korea}
\affiliation{Department of Physics, Korea Advanced Institute of Science and Technology, Daejeon 34141, Republic of Korea}

\author{Youngwook Kim}
\thanks{Corresponding author. y.kim@dgist.ac.kr }
\affiliation{Department of Physics and Chemistry, Daegu Gyeongbuk Institute of Science and Technology (DGIST), Daegu 42988, Republic of Korea}

\date{\today} 

\begin{abstract}
\noindent We demonstrate electrically switchable, non-volatile dipoles in graphene/thin hBN/\texorpdfstring{$\alpha$-RuCl$_3$}{alpha-RuCl3} heterostructures, stabilized purely by interfacial charge transfer across an atomically thin dielectric barrier. This mechanism requires no sliding or twisting to explicitly break inversion symmetry and produces robust ferroelectric-like hysteresis loops that emerge prominently near 30~K. Systematic measurements under strong in-plane and out-of-plane magnetic fields reveal negligible effects on the hysteresis characteristics, confirming that the primary mechanism driving the dipole switching is electrostatic. Our findings establish a distinct and robust route to electrically tunable ferroelectric phenomena in van der Waals heterostructures, 
opening opportunities to explore the interplay between interfacial charge transfer and temperature-tuned barrier crossing of dipole states at the atomic scale.
\end{abstract}

\maketitle

\begin{figure*}[t]
\includegraphics{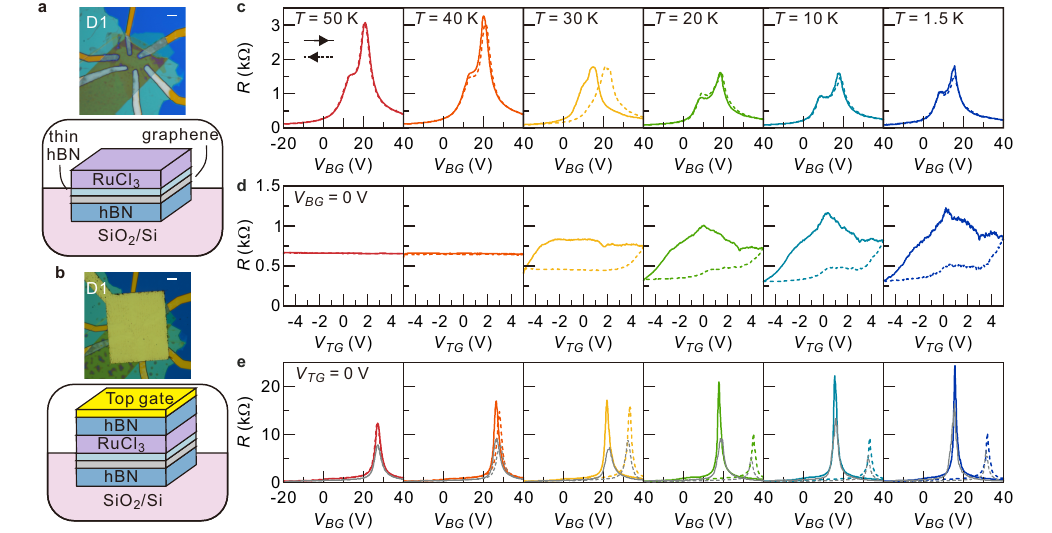} 
\caption{\textbf{Device structure and transport characteristics before and after top-gate fabrication.}(a) Optical image and schematic of a device (D1) without a top gate, used for measurements in panel (c). (b) Optical image and schematic of the same device but with a top gate, used for measurements in panels (d) and (e). Scale bars at panel (a) and (b) for device images are 10~$\mu$m.(c) Four-terminal longitudinal resistance, $R$, measured before top gate fabrication, shown as a function of back gate voltage $V_\mathrm{BG}$ at $B = 0$~T, for temperatures ranging from 1.5~K to 50~K. Solid lines indicate forward sweeps, dashed lines backward sweeps. (d) After top gate fabrication, four-terminal resistance measured as a function of top gate voltage $V_\mathrm{TG}$ at $V_\mathrm{BG} = 0$~V. (e) Resistance as a function of back gate voltage $V_\mathrm{BG}$ measured at a fixed top gate voltage $V_\mathrm{TG} = 0$~V in the same device after top gate fabrication. Grey lines show additional measurements performed after five-month storage under floating top gate conditions. The color coding for temperature is consistent across panels (c)--(e).}
\label{Figure1}
\end{figure*}

\section{Introduction} 

Two-dimensional ferroelectricity, the ability to switch electric polarization in crystals only atoms thick, has emerged as a rapidly growing field~\cite{dawber2005physics, bune1998two, fong2004ferroelectricity, wang2023towards, zhou2022two, wu2021two}. Unlike classical ferroelectrics that rely on collective ionic displacements, polar order in van der Waals systems can arise through less conventional mechanisms, highlighting how the two-dimensional limit, with its strong fluctuations and pronounced interface effects, offers a fundamentally new setting to explore polarization phenomena~\cite{wu2018rise, xiao2018intrinsic, ding2017prediction, yuan2019room, higashitarumizu2020purely, chang2016discovery, gou2023two, chang2020microscopic, Liu2016room, Ghosh2019ultrathin}. One prominent example is sliding ferroelectricity, where stacking two layers in a non-centrosymmetric arrangement creates an out-of-plane dipole that reverses when the layers are laterally shifted~\cite{fei2018ferroelectric, vizner2021interfacial, meng2022sliding, wang2023sliding, niu2022giant, ji2023general, yang2023atypical, Zhang2023visualizing}. Recently, the stacking-induced interface polarization observed in hexagonal boron nitride (hBN) layers illustrates how subtle interlayer misalignments in van der Waals heterostructures can give rise to stable ferroelectric states even without inherent polarity in materials~\cite{yasuda2021stacking, woods2021charge, li2017binary}. 

Combining different two-dimensional materials can lead to interfacial charge transfer that establishes static out-of-plane dipole layers purely through electronic redistribution, without requiring lattice distortions or engineered symmetry breaking~\cite{chen2025ferroelectricity, zheng2020unconventional, Lukas2022Ferroelectricity}. Although such interfacial dipoles have been observed in static measurements, realizing an electrically tunable, switchable dipole response remains largely unaddressed.~\cite{rossi2023direct,rizzo2020charge,rizzo2023polaritonic} In this context, clarifying the precise interplay between interlayer electronic coupling and polar order formation becomes crucial for advancing electrically controlled polarization phenomena at atomic scales.

Graphene paired with $\alpha$-RuCl$_3$, a layered Mott insulator and candidate Kitaev quantum magnet~\cite{plumb2014alpha, banerjee2016proximate, banerjee2017neutron, yokoi2021half, sears2015magnetic}, provides an appealing platform to explore this challenge. Recent studies of graphene/$\alpha$-RuCl$_3$ heterostructures have shown interfacial charge transfer that produces local interfacial polarization at the boundary~\cite{mashhadi2019spin, zhou2019, rizzo2020charge, balgley2022ultrasharp, rizzo2022nanometer, rossi2023direct, kim2023spin}. In the direct-contact limit, however, the large work-function difference drives nearly complete charge transfer, producing a high local carrier density that strongly screens internal electric fields. This pins any interfacial polarization to atomic length scales, preventing a collective response to external gate voltages and thus does not yield stable, switchable dipolar order. Developing heterostructures that moderate the transfer while preserving strong electrostatic coupling, for example by introducing an ultrathin dielectric spacer, remains an important open problem.

To this end, we introduced an ultrathin insulating hBN spacer between graphene and $\alpha$-RuCl$_3$. This few-layer hBN spacer significantly reduces interfacial charge transfer, thereby preventing complete screening, but still permits partial electron redistribution across the interface, stabilizing dipoles that are responsive to gate voltages~\cite{wang2020modulation}. These dipoles manifest as clear ferroelectric-like hysteresis in transport measurements, appearing around 30 K and becoming progressively more stable upon cooling and repeated gating. Drawing on our systematic measurements across temperature, thickness of hBN spacer, and external magnetic fields, we also propose a possible mechanism underlying the observed switchable ferroelectricity.

\section{Results} 
\subsection{Gate-induced interfacial dipole switching and long-term stability} 

To explore this behavior and its dependence on barrier thickness, we fabricated six devices with different interlayer configurations, ranging from a direct graphene/$\alpha$-RuCl$_3$ interface (no hBN, D0) to heterostructures incorporating bilayer (D1, D2), trilayer (D3), seven-layer (D4), and ten-layer hBN spacers (D5). The results presented in Fig.~\ref{Figure1} focus on the device with a bilayer hBN spacer (D1), which serves as a representative case to highlight the essential physics. We performed gate sweep measurements in two stages, before and after top gate fabrication, as shown in Fig.~\ref{Figure1}(a) and ~\ref{Figure1}(b). Figure~\ref{Figure1}(c) presents the longitudinal resistance as a function of back gate voltage $V_\mathrm{BG}$, measured before top gate deposition over temperatures ranging from 1.5~K to 50~K. Solid and dashed lines correspond to forward and backward sweeps, respectively. A hysteresis is observed at $T=30$~K, whereas it is negligible or weak at other temperatures.

After the top gate was fabricated, the gate-dependent behavior revealed several intriguing features. First, as shown in Fig.~\ref{Figure1}(d), sweeping the top gate voltage $V_\mathrm{TG}$ at $V_\mathrm{BG}=0$~V results in relatively modest changes in resistance at 50~K, indicating that the top gate does not strongly modulate the carrier density in the usual way. Second, the hysteresis induced by top gate sweeps, displayed in Fig.~\ref{Figure1}(d), shows a markedly different trend compared to the pre-top-gate measurements. Rather than appearing abruptly at a particular temperature, the hysteresis gradually develops as the temperature decreases, becoming significantly more clear below 30~K. Interestingly, the resistance curves form closed hysteresis loops, unusual in standard graphene-based gating experiments. 

\begin{figure*}
\includegraphics{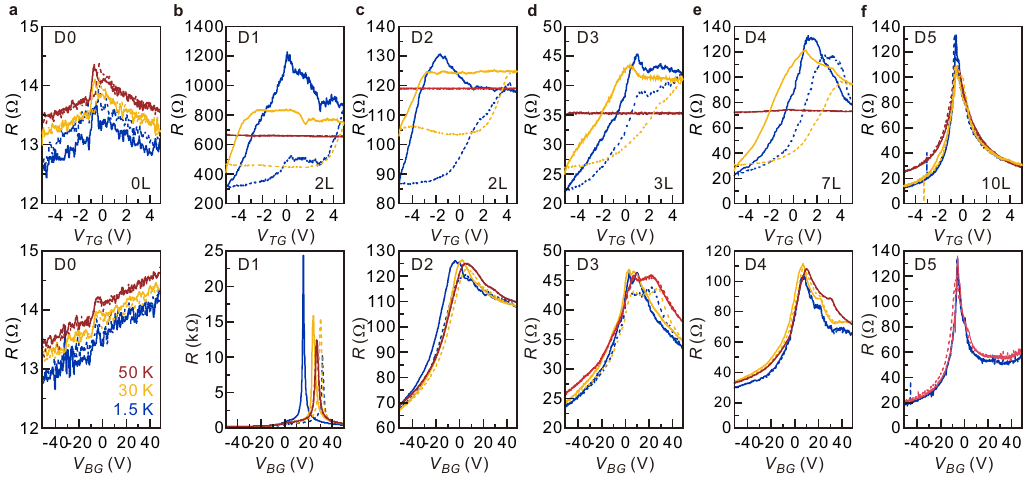}[t]
\caption{\textbf{Spacer-thickness dependence of gate-induced hysteresis in longitudinal resistance.} Dependence of longitudinal resistance on top gate voltage (upper panels) and back gate voltage (lower panels) for (a) device D0 (without thin hBN spacer), (b) device D1 (bilayer hBN), (c) device D2 (bilayer hBN), (d) device D3 (trilayer hBN), (e) device D4 (seven-layer hBN), and (f) device D5 (ten-layer hBN). Device numbers and corresponding hBN thickness are indicated in each upper panel. Red, yellow, and blue curves correspond to measurements at 50~K, 30~K, and 1.5~K, respectively.Device D0, without hBN, shows no measurable hysteresis at any temperature, confirming the critical role of spacer thickness.}
\label{Figure2}
\end{figure*}

Such closed-loop hysteresis strongly resembles ferroelectric switching. In known two-dimensional ferroelectrics, polarization often arises from relative layer sliding or twisting that breaks inversion symmetry~\cite{wang2022interfacial, weston2022interfacial, yang2018origin}. In our geometry, however, the data are consistent with electrostatically driven interfacial dipoles rather than layer sliding or twist-induced symmetry breaking, as indicated by the systematic suppression with spacer thickness [Fig.~\ref{Figure2}]. (Further electrical transport measurements supporting ferroelectric-like switching behavior are presented in Supplementary Note 2 and Supplementary Figure 3.)

These results indicate that a thin hBN spacer plays a crucial role in enabling the observed hysteretic transport features. By introducing the hBN layer, we moderate the charge transfer between graphene and $\alpha$-RuCl$_3$. The spacer is thin enough to enable strong electrostatic coupling across the interface, inducing interfacial polarization (equal and opposite charge accumulation on the facing surfaces) while remaining sufficiently insulating to suppress direct tunneling and avoid immediate electrostatic collapse of the dipole. This balance allows the dipoles to persist and respond collectively to external electric fields. In contrast, when graphene directly contacts $\alpha$-RuCl$_3$ without an intervening dielectric, the large work function mismatch drives excessive charge transfer, leading to strong electrostatic screening. The interface thus becomes effectively overcompensated, preventing the stabilization of dipolar order over appreciable length scales and suppressing ferroelectric-like hysteresis. This interpretation aligns well with previous optical and transport studies of graphene/$\alpha$-RuCl$_3$ and hBN/$\alpha$-RuCl$_3$ heterostructures. These studies inferred interfacial dipoles from phenomena such as modified polariton propagation~\cite{rizzo2020charge, rizzo2023polaritonic}, while devices lacking a spacer fail to exhibit electrical hysteresis~\cite{mashhadi2019spin, kim2023spin}.

A partial alignment of interfacial dipoles may already occur through self-poling during metal deposition or under weak built-in fields associated with the gate stack (see Supplementary Note~3 for a detailed discussion), and subsequent bipolar sweeps of the top gate act as an electrostatic poling process, which is often referred to as training in ferroelectrics~\cite{zheng2010graphene,lee2019ferroelectric}, and further refine this alignment along the out-of-plane field. Once aligned, the dipoles remain uniformly oriented, converting what had been a fluctuating impurity landscape into a static background potential. The practical result is twofold: the hysteresis loops broaden and remain visible well below 30~K, and the graphene channel exhibits a noticeably sharper resistance peak in subsequent back-gate sweeps [Fig.~\ref{Figure1}(e)], which indicates reduced remote-scatterer disorder and an overall cleaner transport environment.

The effectiveness of this electrostatic alignment is governed by the large work-function mismatch between graphene and $\alpha$-RuCl$_3$. With only the back gate available, a clear loop first emerges near 30~K [Fig.~\ref{Figure1}(c)], and at lower temperatures the coercive field exceeds the available gate swing so that the dipoles become rigidly pinned. Incorporating a top gate nearly doubles the accessible vertical field and enables reversible alignment of interfacial dipoles through successive bipolar sweeps, which results in robust, non-volatile switching that persists down to the lowest measured temperatures [Fig.~\ref{Figure1}(d) and (e)]. The recurring involvement of the 30~K window points to an intrinsic interfacial energy scale, which we revisit in the Discussion section.

To further examine the durability of the aligned dipole configuration, device~D1 was stored in an Ar-filled glove box for approximately five months with the top gate left floating (no applied bias). Upon re-cooling, the $R(V_{\mathrm{BG}})$ curves measured under this floating-gate condition [grey lines, Fig.~\ref{Figure1}(e)] remain indistinguishable from those obtained immediately after alignment at $V_{\mathrm{TG}}=0$~V, differing markedly from the pre-alignment response [Fig.~\ref{Figure1}(c)]. Although a rigid horizontal shift of about $+19$~V appears after storage, likely due to slow internal charge redistribution within the encapsulated stack, the sustained hysteresis shapes, coercive voltages, and sharp resistance peaks indicate that the dipole orientation remains stable over a timescale exceeding $10^{6}$~s. This long-term retention, which is far beyond the minute-to-hour decay typical of charge-trapping or ionic-migration processes~\cite{Illarionov2016,Macucci2017}, strongly supports the interpretation that the hysteresis originates from collectively ordered interfacial dipoles rather than extrinsic trapping. (The original, unshifted data before applying the $-19$~V offset are shown in Supplementary Figure 4(a--c).)

The contrasting responses to gating from the top and bottom electrodes, highlighted in Fig.~\ref{Figure1}(d) and (e), further emphasize the role of interfacial dipoles in our system. When the bottom gate is swept, the electric field modulates the carrier density in graphene through the SiO$_2$ and bottom hBN dielectrics without encountering interfacial screening, resulting in conventional gating behavior. In contrast, applying a voltage from the top gate requires the field to penetrate the Mott-insulating $\alpha$-RuCl$_3$ layer and directly engage with the dipole layer at the graphene/hBN/$\alpha$-RuCl$_3$ interface. This leads to strong local screening that suppresses direct field-effect doping, as seen in the modest changes in resistance under top gate sweeps. Instead, the top gate primarily acts by reorienting the interfacial dipoles themselves, giving rise to the ferroelectric-like hysteresis loops we observe. This clear asymmetry between top and bottom gating thus serves as direct evidence for the existence and collective tunability of interfacial dipoles in these heterostructures. Such distinct responses are even more explicitly demonstrated in dual-gate sweep experiments (Supplementary Note 4 and Supplementary Figure 5), highlighting the independent and fundamentally different roles of the top and bottom gates in controlling the dipolar state and carrier density, respectively.

\begin{figure}
\includegraphics{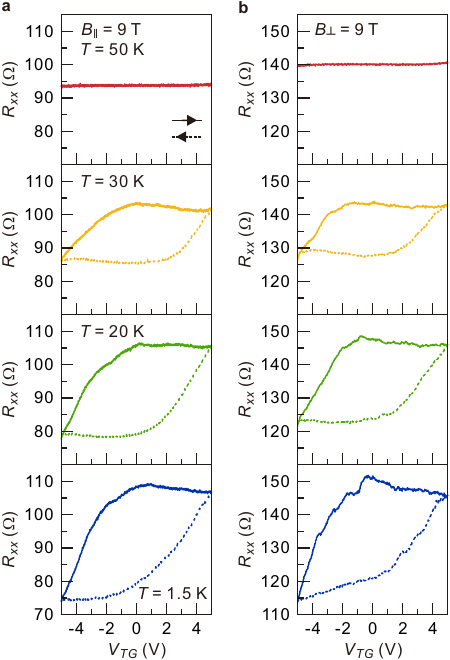}  
\caption{\textbf{Insensitivity of dipole switching to magnetic field orientation.}(a) Four-terminal longitudinal resistance as a function of top gate voltage $V_\mathrm{TG}$ at an in-plane magnetic field $B_{\parallel} = 9$~T and $V_\mathrm{BG} = 0$~V, measured at four temperatures: 50~K, 30~K, 20~K, and 1.5~K for D2. Solid lines represent forward sweeps, and dashed lines represent backward sweeps. (b) Identical measurement as in panel (a), now performed under a perpendicular magnetic field $B_{\perp} = 9$~T.}
\label{Figure3}
\end{figure}

\subsection{Dependence of interfacial dipole switching on hBN spacer thickness} 

Systematic measurements across devices with varying hBN spacer thicknesses, presented in Fig.~\ref{Figure2}, further clarify the role of the spacer in controlling interfacial charge transfer and dipole formation. As the hBN thickness increases from bilayer to ten-layer [Fig.~\ref{Figure2}(b)--(f)], the hysteresis gradually diminishes and eventually vanishes, indicating that thicker hBN layers weaken the interfacial charge redistribution necessary to sustain collective dipole switching. Raman spectroscopy and density functional theory (DFT) calculations (Supplementary Note~5) confirm that interfacial charge transfer decreases with increasing hBN thickness, consistent with this trend. In the device without an hBN spacer (D0, Fig.~\ref{Figure2}(a)), only a very small opening is visible in the top-gate sweep at 1.5~K; this opening shows no systematic temperature dependence and is consistent with extrinsic trapping rather than an appreciable ferroelectric-like response. Compared with the spacer devices, D0 does not display the gate-tunable hysteresis that emerges once a thin spacer is inserted, indicating that the hBN layer plays a key role in enabling a switchable dipolar state. Unlike the bilayer devices (D1, D2), devices with thicker hBN spacers (D3--D5) exhibit their maximum hysteresis near 30~K rather than at 1.5~K; we return to discuss this behaviour in the Thermally Activated Switchable Ferroelectricity section.

\begin{figure*}[t]
\includegraphics{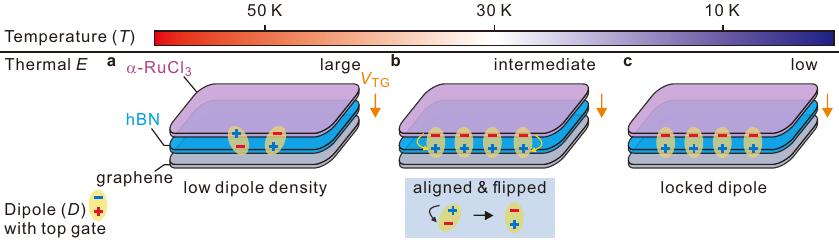}  
\caption{\textbf{Temperature-dependent formation and locking of interfacial dipoles.} Schematic representation of temperature-dependent dipole formation at the graphene/hBN/$\alpha$-RuCl$_3$ heterointerface. The grey, blue, and violet layers represent graphene, hBN, and $\alpha$-RuCl$_3$, respectively. The horizontal bar indicates three characteristic temperature regimes. (a)~At high temperature ($T \sim 50$~K), thermal fluctuations suppress robust dipole formation, resulting in low dipole density. (b)~At intermediate temperature ($T \sim 30$~K), dipoles form and align under an applied gate field due to thermally assisted activation. (c)~At low temperature ($T \sim 10$~K), dipoles are locked and cannot be reversed by the available gate voltage.}
\label{Figure4}
\end{figure*}

\subsection{Magnetic field independence of the ferroelectric-like hysteresis} 

To investigate possible magnetic field effects on the hysteresis around 30~K, we performed systematic gate-dependent measurements on device D2 under strong in-plane ($B_\parallel = 9$~T) and out-of-plane ($B_\perp = 9$~T) magnetic fields, as summarized in Fig.~\ref{Figure3}. The ferroelectric-like hysteresis loops remain nearly unaffected by both magnetic field orientations. Specifically, the hysteresis onset around 30~K, the increase in loop area upon cooling, and the overall shape of the gate response show negligible dependence on magnetic field direction. Additional measurements conducted under continuously varied perpendicular magnetic fields up to 9~T, as well as fixed in-plane fields, further confirm this magnetic-field independence, with minimal changes observed in loop shape and coercive voltage (Supplementary Note~6 and Supplementary Figure 8).

\subsection{Thermally Activated Switchable Ferroelectricity}

The ferroelectricity observed in the device primarily originates at the interface between $\alpha$-RuCl$_3$, hBN, and graphene, where strong electric field gradients arising from charge transfer facilitate the spontaneous formation of dipole moments. Building on this understanding, the observed temperature-dependent switchable ferroelectric behavior can be explained by considering how these interfacial dipoles respond to thermal fluctuations, as illustrated schematically in Fig.~\ref{Figure4}.

First, the two limiting temperature regimes, corresponding to very high and very low temperatures, are relatively straightforward to understand. At temperatures above or near 50~K, thermal fluctuations are strong enough to destabilize the spontaneous electric polarization, thereby preventing the formation of a sufficiently robust ferroelectric state detectable in transport [Fig.~\ref{Figure4}(a)]. Essentially, the free energy gain by developing the polarization is small and the resulting polarization too weak in this temperature range [Fig.~\ref{Figure4}(a)]. In contrast, at lowest temperatures, thermal fluctuations are greatly suppressed, allowing a significant amount of dipoles to appear and align from hBN toward $\alpha$-RuCl$_3$ [Fig.~\ref{Figure4}(c)]. However, in this regime, the dipoles are effectively frozen, as reversing their orientation entails an energy cost that cannot be overcome by the applied gate voltage due to the absence of adequate thermal activation [Fig.~\ref{Figure4}(c)]. Consequently, they are not switchable. This thermally driven scenario provides a consistent explanation for the ferroelectric behavior observed in devices without a top gate [Fig.~\ref{Figure1}(c)]. From the free energy perspective, while the system can significantly lower its energy by developing a polarization, the energy barrier between the two (meta)stable polarization states is too large to be overcome by the applied gate voltage. As already noted in Fig.~\ref{Figure1}, the addition of the top gate and repeated gate sweeps align and stabilize the dipoles, enabling switching down to the lowest temperatures by effectively reducing the nucleation barrier.

The intermediate temperature regime around 30~K represents a balance between two competing effects [Fig.~\ref{Figure4}(b)]: the development of sufficiently strong polarization due to the reduced thermal fluctuations compared to temperatures above 50~K, and the thermal energy being comparable to the energy barrier required to reverse the polarization direction. As a result, in this regime, electric polarization becomes strong enough to be discernible in transport measurements even in the absence of external fields. Simultaneously, the energy barrier between the two polarization states is shallow enough to be overcome by the applied gate voltage, rendering the ferroelectricity switchable. Now we can clearly understand why ferroelectric-like hysteresis consistently appears around 30~K. In this temperature range, the system can significantly lower its free energy by developing a polarization, while the energy barrier between the two dipole states remains sufficiently shallow to be overcome by the applied electric field [Fig.~\ref{Figure4}(b)]. Although the switching behavior is insensitive to external magnetic fields [Fig.~\ref{Figure3}], the relevant energy scale around 30~K is comparable to local excitations in $\alpha$-RuCl$_3$. We do not attribute the switching to magnetic fluctuations; rather, this coincidence indicates that the interfacial environment is particularly responsive to gate-driven dipole reorientation in this temperature window.

Importantly, the disappearance of a measurable hysteresis loop at temperatures above $\sim$50~K does not imply that the electrically trained configuration has been thermally erased. In our picture, the 30~K regime corresponds to an activation window in which thermal fluctuations assist gate-driven transitions between metastable dipolar states. Heating the device in zero field suppresses the collective switching response but does not drive the system back into an unpolarized configuration. Upon cooling, the device returns to the same trained dipolar state, consistent with the long-term stability observed in Fig.~\ref{Figure1}(e).

The temperature range for reversible ferroelectric switching agrees qualitatively with the energy scale associated with the effective switching field, exemplified here by device D1. We evaluate the field from $E_c \approx \xi V_c / t_{\mathrm{eff}}$, using $t_{\mathrm{eff}}\simeq 6.5~\mathrm{nm}$ (20~nm top hBN + 0.8~nm spacer; $\varepsilon_{r}\!\approx\!3.2$), $V_c \simeq 2.0~\mathrm{V}$ at $30~\mathrm{K}$ [Fig.~\ref{Figure1}], and a voltage-partition factor $\xi \simeq 0.03$ (see Methods), which gives $E_c \approx 0.92\times 10^{7}~\mathrm{V\,m^{-1}}$. Taking $p=\eta e d$ with $d=0.5$~nm and a Raman-derived effective charge fraction for D1 of $\eta_{\mathrm{D1}}\approx 0.45$, the dipole–field work becomes
\[
\frac{U}{k_{\mathrm B}} \approx 58~\mathrm{K}\,\eta_{\mathrm{D1}}\left(\frac{E_c}{10^{7}~\mathrm{V\,m^{-1}}}\right)\left(\frac{d}{0.5~\mathrm{nm}}\right)
\approx 24~\mathrm{K},
\]
consistent with the observed onset window near $30~\mathrm{K}$. A similar temperature scale is consistently estimated for other devices (Methods). This consistency substantiates the validity of our understanding of switchable ferroelectricity driven by thermal fluctuations.

Furthermore, this thermal activation effect accounts for the ferroelectric behaviors in devices with thicker hBN spacers [Fig.~\ref{Figure2}(d--f)], where reduced charge transfer results in a weaker dipolar response. Despite the smaller polarization, these heterostructures still display a clear hysteresis peak near 30~K that diminishes substantially upon cooling. Around 30~K, thermal energy enables collective barrier crossing even under the reduced built-in polarization and weaker gate coupling. At lower temperatures, the required switching field exceeds the available gate range as thermal energy decreases, pinning the dipoles and suppressing switching.

This thickness-dependent contrast between bilayer and thicker-spacer devices can be understood as a balance between dipole strength and the reorientation barrier. For bilayer hBN (D1, D2), stronger interfacial charge redistribution yields larger dipole moments and lower barriers, which keeps the loops large and switchable down to 1.5~K after field alignment. In contrast, thicker spacers (D3--D5) weaken the interfacial coupling and increase the effective barrier, shifting the maximal loop size to the thermally assisted regime near 30~K.

Another possible microscopic contributions of the observed switchable ferroelectricity include slight Ru--Ru bond distortions and octahedral tilts previously visualized at the atomic scale~\cite{ziatdinov2016atomic}, as well as the recently discovered incommensurate charge super-modulation hosting hidden dipole order in $\alpha$-RuCl$_3$~\cite{zheng2024incommensurate}. While residual strain or stacking faults introduced during fabrication may contribute to the development of ferroelectricity in our device, such effects are typically random and localized, making them unlikely to account for the device-scale hysteresis observed.

\section{Discussion} 

Our results demonstrate that interfacial charge transfer creates a stable and gate-tunable dipolar landscape in graphene/hBN/$\alpha$-RuCl$_3$ heterostructures. Near 30~K, the interplay between thermal fluctuations and electrostatic energy barriers provides optimal conditions for electrically driven dipole switching, resulting in clear hysteresis. At higher temperatures, strong thermal fluctuations disrupt the formation of coherent dipolar ordering, whereas at lower temperatures, the increased coercive field pins the dipoles, suppressing their switchability. This temperature-dependent balance between thermal activation, interfacial charge redistribution, and gate-driven electrostatic control provides a physical explanation for the observed dipole switching, highlighting the potential of engineered heterostructures to achieve tunable ferroelectric-like behavior in atomically thin materials.

- \textit{Note added}: During the preparation of this manuscript, we became aware of a related independent study~\cite{liu2025dynamic} reporting interfacial dipole formation in graphene/hBN/$\alpha$-RuCl$_3$ heterostructures. 

\section{Methods} 
\subsection*{Device fabrication}
Device fabrication involved two distinct approaches, differentiated by device configuration. Fabrication of Device 1 starts with mechanically exfoliating monolayer graphene onto a SiO\textsubscript{2}/Si substrate and patterning it into a Hall-bar geometry using local anodic oxidation via atomic force microscopy. A top hBN flake ($\sim$20 nm thick) was mechanically exfoliated onto a Gel-Pak substrate, while a thin hBN layer was exfoliated onto a separate SiO\textsubscript{2}/Si substrate. The graphene and thin hBN layers were sequentially picked up using the top hBN flake as a stamp. To place the thin hBN spacer above the graphene layer, we employed a full dry flipping process using two Gel-Pak films, allowing clean reversal of the stacking order. The flipped stack was subsequently transferred onto a SiO\textsubscript{2}/Si substrate. The Gel-Pak film used as a carrier of the flipped stack was treated with UV/Ozone (185 nm, 15 mW cm\textsuperscript{-2}) for 3 minutes to improve surface adhesion.~\cite{kim2024full}.

A 950 K PMMA resist was spin-coated at 6000 rpm and used for electrode patterning via electron-beam lithography. The exposed areas were etched using sequential CF\textsubscript{4} plasma (40 sccm, 40 W) and O\textsubscript{2} plasma (50 sccm, 100 W) under a chamber pressure of 3 $\times$ 10\textsuperscript{-3} mbar. Electrodes were deposited by thermal evaporation of 4 nm Cr followed by 40 nm Au at a base pressure of 2 $\times$ 10\textsuperscript{-7} mbar. After lift-off, residual contamination was removed by contact-mode atomic force microscopy cleaning, applying a tip force of 5–30 nN. The $\alpha$-RuCl\textsubscript{3} flake was exfoliated onto a separate Gel-Pak substrate and transferred onto the prepared heterostructure. To improve interface cleanliness, contact-mode atomic force microscopy cleaning was also applied after the $\alpha$-RuCl\textsubscript{3} transfer. Finally, to prevent chemical degradation of $\alpha$-RuCl\textsubscript{3}, the entire structure was encapsulated with an additional top hBN layer. A top metal gate with 4 nm Cr and 100 nm Au was fabricated on the encapsulated stack by thermal evaporation. A schematic illustration of the fabrication process is provided in Supplementary Figure 1.

For the remaining devices, a conventional bottom-electrode method was employed. Electrodes were first patterned on an hBN flake exfoliated onto a SiO\textsubscript{2}/Si substrate using e-beam lithography, followed by CF\textsubscript{4} plasma etching under the same conditions as for Device 1. Cr/Au electrodes were then deposited with a total thickness matching that of the bottom hBN layer to minimize mechanical strain on the heterostructure after releasing. Contact-mode atomic force microscopy cleaning was applied after lift-off to remove residue. The heterostructure, composed of hBN/$\alpha$-RuCl\textsubscript{3}/thin hBN/graphene, was picked up using an Elvacite stamp~\cite{mashhadi2019spin, kim2023spin, kim2019even, kim2022robust, kim2023orbitally}, which was chemically removed after transfer. Each device was finally completed by fabricating a top metal gate using the same procedure as for Device 1.

\subsection*{Training procedure}
Before collecting the gate-dependent data presented in the main text and Supplementary Information, the interfacial dipolar state in the graphene/hBN/RuCl$_3$ stack was initialized through the following steps:
\begin{enumerate}
    \item After completing the top-gate fabrication and wiring, the device was cooled from room temperature down to 1.5\,K with $V_{\mathrm{TG}} = 0$\,V and $V_{\mathrm{g}} = 0$\,V.

    \item At 1.5\,K, we applied the first bipolar top-gate cycle. Specifically, $V_{\mathrm{TG}}$ was ramped from 0\,V to $-5$\,V, then swept from $-5$\,V to $+5$\,V (forward sweep), immediately swept back from $+5$\,V to $-5$\,V (backward sweep), and finally returned to 0\,V.

    \item The device was then warmed, and at several temperatures during the first warm-up, the same bipolar $V_{\mathrm{TG}}$ cycle was repeated: from 0\,V to $-5$\,V, then $-5$\,V to $+5$\,V, then $+5$\,V to $-5$\,V, and back to 0\,V. These repeated cycles progressively stabilized the dipolar configuration.

    \item After completing this sequence, $V_{\mathrm{TG}}$ was fixed at 0\,V and the device was cooled back down to 1.5\,K.

    \item Following this training, all back-gate measurements were performed with $V_{\mathrm{TG}} = 0$\,V. At each measurement temperature, $V_{\mathrm{g}}$ was swept by ramping from 0\,V to $-50$\,V, then sweeping from $-50$\,V to $+50$\,V (forward), sweeping back from $+50$\,V to $-50$\,V (backward), and returning to 0\,V. The data shown in Fig.~1, Fig.~2, Fig.~3, and Supplementary Figure~3 and~4 were acquired in this manner while warming the trained device.
\end{enumerate}

\subsection*{Transport measurements}
Electrical transport measurements were performed using low-frequency lock-in techniques in a four-terminal configuration, with an AC current of 100~nA applied at a frequency of 13.333~Hz. All measurements were conducted in cryogenic environments with a base temperature of 1.4~K. Out-of-plane magnetic field measurements were performed using a Janis DryMag system (up to 9~T), while in-plane field measurements were conducted using an Oxford Teslatron system (up to 12~T). Gate-dependent resistance was measured as a function of magnetic field (0--9~T), temperature (1.5--50~K), and top gate voltages. For D~1, gate sweep measurements were performed both before and after top gate fabrication.

\subsection*{Theoretical model and parameter extraction}
\paragraph*{Model.}
We estimate the switching energy scale from the dipole–field coupling
\(U \simeq p\,E_c\) with \(p=\eta e d\) and \(E_c \simeq \xi V_c/t_{\mathrm{eff}}\).
Here \(d\) is the atomic-scale interfacial separation (we use \(d=0.5~\mathrm{nm}\)),
\(\eta\) is a dimensionless effective charge fraction, \(V_c\) is the switching voltage,
\(t_{\mathrm{eff}}\) is the effective dielectric thickness between the top gate and the interface,
and \(\xi\) (\(0<\xi<1\)) accounts for voltage partition.

\paragraph*{Extraction of \(V_c\).}
From forward/backward \(R\)–\(V_{\mathrm{tg}}\) sweeps at \(T\approx 30~\mathrm{K}\),
we identify switching points (extrema of \(\partial R/\partial V_{\mathrm{tg}}\)),
define \(\Delta V_{\mathrm{hys}}\) as their separation, and set
\(V_c=\Delta V_{\mathrm{hys}}/2 \simeq 1\text{--}2~\mathrm{V}\) (typical \(\Delta V_{\mathrm{hys}}=2\text{--}4~\mathrm{V}\)).

\paragraph*{Effective thickness \(t_{\mathrm{eff}}\) and partition factor \(\xi\).}
For the stack \(\mathrm{Top\;gate}/\mathrm{hBN}_{\mathrm{top}}/\alpha\text{-}\mathrm{RuCl_3}/\mathrm{hBN}_{\mathrm{thin}}\),
\[
t_{\mathrm{eff}}=\frac{t_{\mathrm{top}}}{\varepsilon_{r,\mathrm{hBN}}}+\frac{t_{\mathrm{thin}}}{\varepsilon_{r,\mathrm{hBN}}},
\quad \varepsilon_{r,\mathrm{hBN}}\approx 3.2.
\]
Using \(t_{\mathrm{top}}\approx 20~\mathrm{nm}\) and \(t_{\mathrm{thin}}=0.8,\,1.6,\,3.8,\,5.4~\mathrm{nm}\)
(2L, 3L, 7L, 10L) gives \(t_{\mathrm{eff}}\approx 6.5,\,6.75,\,7.44,\,7.94~\mathrm{nm}\) at 30~K.
We obtain \(\xi\) from a series-capacitor partition between the top hBN, the spacer hBN, and the \(\alpha\)-RuCl\(_3\) layer:
\[
\xi \;=\; \frac{C_{\mathrm{top}}}{C_{\mathrm{top}}+C_{\mathrm{eq}}}\times
\frac{C_{\mathrm{Ru}}}{C_{\mathrm{sp}}+C_{\mathrm{Ru}}},\quad
C_{\mathrm{eq}}=\frac{C_{\mathrm{sp}}\,C_{\mathrm{Ru}}}{C_{\mathrm{sp}}+C_{\mathrm{Ru}}},
\]
with \(C_{\mathrm{top}}=\varepsilon_0\varepsilon_{r,\mathrm{hBN}}/t_{\mathrm{top}}\),
\(C_{\mathrm{sp}}=\varepsilon_0\varepsilon_{r,\mathrm{hBN}}/t_{\mathrm{thin}}\),
and \(C_{\mathrm{Ru}}=\varepsilon_0\varepsilon_{\perp,\mathrm{RuCl_3}}/t_{\mathrm{Ru}}\).
Using \(\varepsilon_{\perp,\mathrm{RuCl_3}}\approx 5\text{--}6\) and \(t_{\mathrm{Ru}}\sim 10~\mathrm{nm}\) gives representative
\(\xi\) values at 30~K of
\(\xi_{\mathrm{2L}}\approx 0.03\), \(\xi_{\mathrm{3L}}\approx 0.058\), and \(\xi_{\mathrm{7L}}\approx 0.128\).
(We do not include the \(\alpha\)-RuCl\(_3\) thickness in \(t_{\mathrm{eff}}\); its screening only enters through \(\xi\).)

\paragraph*{Raman-based estimate of \(\eta\).}
Graphene Raman features \((\omega_G,\omega_{2D},\Gamma_G)\) at zero bias are converted to a carrier density
\(\Delta n\) via our calibration; typical values are \(\Delta n \sim (5\text{--}15)\times10^{12}\,\mathrm{cm^{-2}}\).
The associated bound charge density is \(\sigma=e\,\Delta n\), yielding an areal interfacial polarization
\(\Pi=\sigma d\). To remain consistent with the parameterization \(p=\eta e d\), we report the dimensionless
factor \(\eta\) that reproduces this polarization scale; across devices we obtain \(\eta\simeq 0.2\text{--}0.4\).
For the device-specific estimates below we use Raman-guided representative values
\(\eta_{\mathrm{D1}}=\eta_{\mathrm{D2}}\approx 0.45\) (bilayer spacers, \(\Delta n\approx 2.0\times10^{13}\,\mathrm{cm^{-2}}\)),
\(\eta_{\mathrm{D3}}\approx 0.225\) (\(\Delta n\approx 1.0\times10^{13}\,\mathrm{cm^{-2}}\)),
and \(\eta_{\mathrm{D4}}\approx 0.146\) (\(\Delta n\approx 0.65\times10^{13}\,\mathrm{cm^{-2}}\)).

\paragraph*{Resulting field and energy scales (devices D1–D4).}
With \(d=0.5~\mathrm{nm}\) and the above parameters, the effective fields
\(E_c=\xi V_c/t_{\mathrm{eff}}\) and the characteristic scales \(U/k_{\mathrm B}=58~\mathrm{K}\,\eta\,(E_c/10^{7}\,\mathrm{V\,m^{-1}})\) at 30~K are:
\begin{itemize}
\item \textbf{D1, D2 (2L)}: \(t_{\mathrm{eff}}\approx 6.5~\mathrm{nm}\), \(\xi\approx 0.03\), \(V_c\simeq 2.0~\mathrm{V}\) \(\Rightarrow\)
\(E_c\approx 0.92\times 10^{7}~\mathrm{V\,m^{-1}}\), \(\eta_{\mathrm{D1}}\approx 0.45\) \(\Rightarrow\)
\(\boxed{U/k_{\mathrm B}\approx 24.1~\mathrm{K}}\).
\item \textbf{D3 (3L)}: \(t_{\mathrm{eff}}\approx 6.75~\mathrm{nm}\), \(\xi\approx 0.058\), \(V_c\simeq 2.0~\mathrm{V}\) \(\Rightarrow\)
\(E_c\approx 1.72\times 10^{7}~\mathrm{V\,m^{-1}}\), \(\eta_{\mathrm{D3}}\approx 0.225\) \(\Rightarrow\)
\(\boxed{U/k_{\mathrm B}\approx 22.4~\mathrm{K}}\).
\item \textbf{D4 (7L)}: \(t_{\mathrm{eff}}\approx 7.44~\mathrm{nm}\), \(\xi\approx 0.128\), \(V_c\simeq 1.5~\mathrm{V}\) \(\Rightarrow\)
\(E_c\approx 2.58\times 10^{7}~\mathrm{V\,m^{-1}}\), \(\eta_{\mathrm{D4}}\approx 0.146\) \(\Rightarrow\)
\(\boxed{U/k_{\mathrm B}\approx 21.9~\mathrm{K}}\).
\end{itemize}
These values provide a conservative, single-dipole upper bound and are consistent with the reversible-switching window observed near \(30~\mathrm{K}\),
bearing in mind that nucleation-limited dynamics can shift the actual onset by an \(O(1)\) factor.

\subsection*{Raman spectroscopy}
Raman spectroscopy was performed at room temperature (300~K) using a 532~nm excitation laser and a 600~g/mm grating. The laser power was set to 0.94~mW for all configurations, except for the device with a monolayer hBN spacer, for which 1~mW was used. The integration time was fixed at 60~s. All measurements were conducted in circular polarization resolved geometry, using left-circularly polarized incident light and right-circularly polarized detection (LR configuration), and vice versa (RL configuration).

\subsection*{DFT calculation}
First-principles calculations based on DFT were performed using the projector augmented wave (PAW) method as implemented in the Vienna Ab Initio Simulation Package (VASP)~\cite{kresse1996efficient, perdew1996generalized}. The Perdew–Burke–Ernzerhof (PBE) form of the generalized gradient approximation (GGA) was employed for the exchange–correlation functional. A plane-wave energy cutoff of 400 eV was used throughout all calculations. Brillouin zone integration was carried out using a 5$\times$5$\times$1 Monkhorst–Pack k-point mesh for the graphene/hBN/$\alpha$-RuCl$_3$ heterostructure. The heterostructure was constructed by stacking a 5$\times$5 graphene supercell, a 5$\times$5 hBN supercell (compressed uniformly by 1.7$\%$ to match the graphene lattice), and a 2$\times$2 $\alpha$-RuCl$_3$ supercell (expanded uniformly by 1.5$\%$ to fit the stack). To accurately capture the correlated nature of Ru 4d orbitals, an effective Hubbard U correction (U$_{eff}$ = 2 eV) was applied following the Dudarev approach~\cite{dudarev1998electron}. All atomic positions were fully relaxed until the residual forces were less than 0.02 eV/\AA, with spin–orbit coupling (SOC) incorporated self-consistently. Van der Waals interactions were taken into account via the Tkatchenko–Scheffler (DFT-TS) dispersion correction scheme~\cite{bucko2013improved}. A vacuum spacing of 18 \AA was introduced along the out-of-plane direction to eliminate spurious interlayer interactions. For the magnetic configuration of $\alpha$-RuCl$_3$, a zigzag antiferromagnetic ordering of Ru atoms was assumed and optimized within the GGA framework.

\section{Data availability} 
Source data for all main and supplementary text figures are provided with this paper and are also available via the Figshare repository at https://doi.org/10.6084/m9.figshare.30747287

\section{References}

\section{Acknowledgment} 
We thank Erik Henriksen for helpful discussions.The work from DGIST was supported by the National Research Foundation of Korea (NRF) (Grant No. RS-2025-00557717, RS-2023-00274875, RS-2023-00269616)  and the Nano and Material Technology Development Program through the National Research Foundation of Korea (NRF) funded by Ministry of Science and ICT (No. RS-2024-00444725). We also acknowledge the partner group program of the Max Planck Society. G.Y.C. is financially supported by Samsung Science and Technology Foundation under Project Number SSTF-BA2401-03, the NRF of Korea (Grants No. RS-2023-00208291, RS-2024-00410027, 2023M3K5A1094810, RS-2023-NR119931, RS-2024-00444725, RS-2023-00256050, IRS-2025-25453111) funded by the Korean Government (MSIT), the Air Force Office of Scientific Research under Award No. FA23862514026, and Institute of Basic Science under project code IBS-R014-D1. This work was performed in part at Aspen Center for Physics, which is supported by National Science Foundation grant PHY-2210452. K.-H. J was supported by Global-Learning and Academic research institution for Master’s·PhD students, and Postdocs (LAMP) Program of the National Research Foundation of Korea (NRF) grant funded by the Ministry of Education (No. RS-2024-00443714). The works at POSTECH were supported by National Research Foundation of Korea (No. RS-2024-00410027 and No. 2022M3H4A1A04074153). K.W. and T.T. acknowledge support from the JSPS KAKENHI (Grant Numbers 21H05233 and 23H02052) , the CREST (JPMJCR24A5), JST and World Premier International Research Center Initiative (WPI), MEXT, Japan.

\section{Author contributions} 
S. K, Y. K and G.Y.C conceived the project. S. K, J. C and D. K carried out the device fabrication and performed the low-temperature measurement with Y. K. and J. F. The theory was performed by K.-H. J and G. Y. C., J. H. Y and J. S .K conducted Raman Spectroscopy. T. T and K. W synthesized the h-BN crystals. All authors contributed to the manuscript writing. 

\section{Competing interests} 
The authors declare no competing interest
\end{document}